# On-the-fly Segmentation Approaches for X-ray Diffraction Datasets for Metallic Glasses

Fang Ren[1, *, †], Travis Williams[2, †], Jason Hattrick-Simpers[3], Apurva Mehta[1,*]

[1]Stanford Synchrotron Radiation Lightsource, SLAC National Accelerator Laboratory, Menlo Park, CA 94025, USA. [2]College of Engineering and Computing, University of South Carolina, Columbia, SC 29208, USA. [3]Materials for Energy and Sustainable Development Group, National Institute of Standards and Technology, MD 20899, USA

* e-mail: fangren@slac.stanford.edu, mehta@slac.stanford.edu

[†]These authors contributed equally to this work.

**ABSTRACT (100 words)**

Investment in brighter sources and larger detectors has resulted in an explosive rise in the data collected at synchrotron facilities. Currently, human experts extract scientific information from these data, but they cannot keep pace with the rate of data collection. Here, we present three on-the-fly approaches - attribute extraction, nearest-neighbor distance, and cluster analysis - to quickly segment x-ray diffraction (XRD) data into groups with similar XRD profiles. An expert can then analyze representative spectra from each group in detail with much reduced time, but without loss of scientific insights. On-the-fly segmentation would, therefore, result in accelerated scientific productivity.

**KEYWORDS: X-ray diffraction (XRD), amorphous metal**

# INTRODUCTION

Over the past 3 decades, investment in brighter sources and the development of larger and faster area detectors has resulted in an explosive rise in the amount of data collected at synchrotron facilities. The rate of new discoveries, however, has not yet kept pace with the rate of data collection, mostly because data is still curated and analyzed by humans. Though humans are superb at utilizing domain knowledge to discover new materials and to develop new theories, they are too slow to keep pace with the accelerated pace of data acquisition. New automated methods are needed to extract trends and patterns from data in nearly real time to facilitate expert analysis.

One technology that generates large scale data via synchrotron lightsources and is critically in need of automatic data analysis tools is high-throughput x-ray diffraction (HiTp XRD)[1-5]. HiTp experiments typically involve mapping the impact of composition and processing on the structure of a material using a combinatorial sample (combi library) and other rapid synthesis methods[6-8]. HiTp is regularly used to identify and optimize novel materials for a number of important technologies. For instance, two classes of metal alloy systems, namely metallic glasses (MG) and high entropy alloys (HEA), are emerging as technologically advanced materials to be used as high temperature and high hardness coatings[9-13]. These materials are composed of three[14] or more different elements[15], and processing conditions play a significant role in synthesizing them. The composition-processing search space is vast, and trial and error search through traditional serial experiments is often fruitless. HiTp XRD studies of these materials are advantageous as both systems are only kinetically or entropically stabilized, and physiochemical theories[16-19] are not yet very reliable and can yield discrepancies when predicting a particular MG or HEA system.

In order to develop an on-the-fly data analysis routine for technologies like HiTp XRD, a balance is sought between domain knowledge and computer resources. It is advised to avoid shipping large amount of raw data across the network, therefore on-the-fly data analysis should occur on a commodity computer available at any data collection end station. This requires that the computing requirement for analysis must be affordable by a commodity computer, which can be achieved by incorporating domain knowledge. On the other hand, because the data collection end stations run 24 hours per day, constant human inputs should be avoided. The first goal of this kind of data analysis is to develop automated capabilities for rapidly transforming raw XRD data emerging from experiments into scientifically meaningful formats. The second goal is to extract, aggregate and disseminate high-level information in computable form and organize them into a database, which could be utilized to improve data quality and to ensure sufficient data have been collected from a sample in nearly real time. The third goal is to rapidly segregate data into groups where the XRD patterns are similar inside each group, so that expert knowledge only needs to be applied to a few representatives chosen from each group. There has been good progress on automating the first two goals, and those are summarized in our recent paper[20]. In this paper, we used a HiTp XRD dataset taken for the Co-Fe-Zr ternary system, a potential MG material, as an example to explore approaches for the third goal.

To fulfill the third goal, we need automated approaches that require little human intervention to rapidly segment the dataset as the data are collected. As soon as the experimental cycle is complete, scientists will have access to data that are already segmented, so that they can spend less time on routine data analysis tasks and focus on expert analysis. We have adopted three approaches which comprise this strategy. In the first approach, we extracted attributes with direct linkage to the classification of interest, for example, the width, intensity, and position of diffraction peak to the degree of crystallinity. This approach works on one diffraction spectrum at a time. In the second approach, we took advantage of the domain knowledge that within a phase region, data points closer to each other in physical space ("neighbors") are more likely to be similar than those at the phase boundary. This approach works on two pairs of diffraction spectra at a time. In the third approach, we applied several well-established unsupervised clustering methods to the whole dataset (more than 1200 spectra for Co-Fe-Zr dataset).

**RESULTS AND DISCUSSIONS**

<u>Attribute extraction</u>

In the first approach, attributes are extracted from a single spectrum and visualized in elemental space to help understand the property change with the compositions. Attributes are defined as the quantitative scientific information that can be extracted from XRD data, including metadata. Metadata store information of experimental conditions, for example, x-ray beam flux, temperature, and motor positions. From each XRD pattern (either a 2D image or a 1D spectrum), several attributes can be extracted, and each of them represents a specific material property. In our recent publication[20], we discussed several attributes that can be used to assess the global data quality, for example, crystallinity (maximum intensity ($I_{max}$) divided by average intensity ($I_{ave}$)), crystalline texture, and number of peaks. These attributes can also be used for visually segmenting the dataset. In this paper, we will introduce three additional attributes: the position of the first sharp diffraction peak (FSDP), the full width at half maximum (FWHM) of FSDP, and its intensity, extracted from XRD 1D spectra.

The FWHM, intensity, and position of FSDP can be identified through fitting an XRD 1D spectrum, following proper background subtraction. However, background subtraction for XRD patterns from a combi library can be challenging. Because the combi library usually covers a broad range of materials, crystalline and amorphous, multi- and single-phased, the XRD spectra may exhibit very different profiles. Scattering from the substrate and other contributions to the background can cause further difficulties. The background subtraction method also needs to be fast enough for real-time processing and can be applied to all of the spectra in a combi library without human intervention. We have chosen a multi-step method for background subtraction, and the details are included in the Supplementary Information.

Peak fitting can then be applied to the background subtracted spectra. Based on the shape of XRD peaks, Voigt function, a convolution of Gaussian and Lorentzian functions, is often chosen to describe the shape of the XRD peaks. However, because of its high computation cost, Gaussian-Lorentzian sum (GLS) function and Gaussian-Lorentzian product (GLP) function are often used instead[21]. In this paper, GLS will be used for fitting the XRD peaks to obtain the values for FWHM, peak position, and peak intensity. FWHM of FSDP is often used to estimate the ordered domain size in the sample, through Scherrer function[22]. Position of FSDP can be used as a measure of the unit cell volume and to estimate material density. The intensity of FSDP can serve as a sanity check for the crystallinity ($I_{max}/I_{ave}$) since the maximum intensity of a spectrum can sometimes be from the background, which could falsely label a sample as being crystalline. The three attributes, FWHM, position, and intensity of FSDP, together with the other three attributes, crystallinity ($I_{max}/I_{ave}$), texture, and number of peaks that were discussed in our recent publication are plotted in FIG. 1 in the elemental space.

Because each of the attributes represent only one physical property, the segmentation of the dataset should be performed by taking into consideration all of the attributes. For example in FIG. **1**, the sample is crystalline (FIG. **1**a and e) but not textured (FIG. **1**b) around Co-Fe binary region. If we only use texture attribute to segment the data, we may assume the samples around Co-Fe binary region are the similar with those in the center of the ternary. Another example is that the peak position map (FIG. **1**f) indicates that the spectra change gradually with the Zr concentration, which is less obvious from other attribute maps. Although it did not occur in the current library, new peaks appearing at markedly different positions in space would provide sharp visible boundaries in such mappings. By considering the six attributes, we can conclude that a potential phase boundary appears in the Zr-rich region (Zr > 60 at.%), as indicated by FIG. **1**a, b, c and e. Another possible phase boundary lies close to Co-Fe binary line (Zr < 15 at.%), as indicated by FIG. 1a, c and e. From the gradient seen in FIG. **1**f, we conclude that Zr concentration is a major contributor to the shift of FSDP positions.

Nearest-neighbor distance (NND)

NND approach was introduced first in our recent publication[20]. There are two major differences between this approach and attribute mapping. First, no attribute needs to be defined and extracted from the spectra in the NND approach; instead, the whole intensity array is used as raw features. Therefore, NND approach is less reliant on domain knowledge. Second, the NND approach, as its name indicates, uses not only the spectrum-of-interest, but also its nearest two neighbors in the upper stream of the measurements. One important assumption for NND approach is that XRD patterns change more dramatically (having larger distances between neighbors) across a phase boundary than within a phase region. NND often performs as well as sophisticated clustering methods even though it uses only 3 XRD patterns at a time, and therefore it requires much reduced memory usage and the computation cost and is significantly faster and cheaper than cluster analysis. This is especially true for large datasets. Computational resources needed for on-the-fly NND analysis of a large dataset are the same as that for a smaller dataset, whereas computational resources needed for cluster analysis scale non-linearly with the size of the dataset. Because of its low requirement on computer resources, we performed NND analysis on both 1D XRD spectra (with n features) and 2D XRD image (with n*m features for a diffraction pattern with n*m pixels) for distance mapping on a generic desktop machine in real-time. Because 2D images contain much more information than 1D spectra, such as crystalline texture, a more comprehensive analysis can be performed by using 2D images with only moderate additional computation cost. Taking the dataset shown here as an example, it took 0.03 s to calculate the neighbor distances for each 1D spectrum and 6.9 s for each 2D image on a generic desktop, while it took 30 s to collect one diffraction image. Another advantage for NND analysis is that adjacent data points on the same combi wafer share very similar backgrounds unless the wafer substrate has abrupt defects, so background subtraction is no longer a pre-requisite for NND analysis. In this study, because we used three combi wafers to cover one ternary system in order to achieve a slower composition gradient, some variances between the three wafer substrates are unavoidable. For NND analysis, these variances do not need to be taken into account. The main constraint for using this approach is that, in order to make a comparison of spectra between nearest neighbors, the scanning grids and order need to be pre-defined to ensure the physical distances between the nearest neighbors are consistent, so that the spectra are compared with their physical distances fixed. The sampling points also need to be near each other on a combi library to monitor local changes on the wafer. If the samples were chosen randomly from a wafer, the physical distances between neighbors may vary, and this approach cannot be applied anymore. A clustering method needs to be used in this case.

Cosine dissimilarity matrix is chosen to characterize the distances. The advantages of cosine distance matrix have been discussed in a recent paper[23]. Another advantage of cosine distance matrix is that it saves the efforts to normalize the XRD intensity caused by variations in incident x-ray intensity or differences in exposure time, because cosine function only measures the angle between two vectors (spectra) in feature space, the change in absolute intensity will not impact the cosine dissimilarities. The cosine nearest-neighbor distance maps of Co-Fe-Zr ternary using both 1D spectra and 2D images are shown in the inset of FIG. **2**. From FIG. **2**, a potential phase boundary is observed at the Zr-rich region and along the Co-Fe binary line, which is consistent with attribute mapping in FIG. **1**. We were also interested in exploring whether 2D distance maps provide more information; therefore, the distributions of the neighbor distances are plotted using histogram to elucidate the comparison between 1D and 2D NND analysis. From the histograms in FIG. **2**, both distributions are extremely right skewed, because the data points within single phases that have small distances from neighbors are in majority, and the data points on the phase boundaries with large distances from neighbors are in minority. In other words, most of the XRD patterns are similar to their neighbors. The 2D image NND histogram appears to be multi-modal and more skewed compared to the 1D spectrum NND mapping, indicative of enhanced differences between neighbors for 2D images. The enhanced contrast in 2D NND is evident in the splitting of the low difference distribution and the tail at the high difference, highlighted by the two circled regions in FIG. 2. The first circle around -7.8 is due to the texture variation of the diffraction patterns from silica substrate, while the second circle from -5 to -2 is due to the rich crystalline texture in Zr-rich region (Zr > 80 at.%), where NND map of 2D images provide a higher contrast than its 1D spectrum counterpart, as shown in the magnified view in FIG. 2. We expect that for a ternary with richer texture information, the advantage of using 2D images for NND mapping will become even more obvious.

Cluster analysis

Cluster analysis has been adopted in many fields to segment large data into groups to explore trends and patterns hidden in large multi-dimensional data. There has been much progress in developing advanced phase mapping algorithms[24-27] for multi-phased crystalline materials. However, because this article focuses on MG and HEA systems whose XRD patterns contain much fewer numbers of peaks and much wider range of peak shapes (both amorphous and crystalline peaks), these characteristics often challenge many of the clustering approaches developed for crystalline and untextured materials. Four common clustering analysis methods[28] have been chosen: k-medoids, Density-based spatial clustering of applications with noise (DBSCAN), agglomerative clustering (Hierarchical clustering), and spectral clustering. These four methods were chosen among many others because they can take pre-computed dissimilarity matrix (cosine dissimilarity is used in this case) and are scalable with large samples. K-medoids algorithm tries to minimize the distance between points labeled to be in a cluster and a point designated as the center of that cluster. In contrast to a similar and also commonly used clustering method, k-means, k-medoids works on any arbitrary dissimilarity matrices. DBSCAN algorithm treats the data as areas of high density separated by areas of low density, and thus it does not assume any particular cluster shape. The parameters for DBSCAN include the maximum distance between the core sample and its neighbors in the same cluster, and the minimum sample numbers within a cluster. The spectral clustering performs a dimensionality reduction followed by k-means in low dimensional space. The agglomerative clustering performs a hierarchical clustering using a bottom-up approach: each data point forms its own cluster, and then similar clusters are merged together. This step is repeated until the number of clusters is the same as specified. K-medoids, spectral clustering, and agglomerative clustering require the number of clusters to be specified.

For the clustering methods that the number of clusters needs to be specified, cluster numbers ranging from 2 to 10 are examined, and the best results are shown here. K-medoids clustering (FIG. 3a) seems to be affected by background significantly even after automatic background subtraction, and this method runs much slower than other methods. DBSCAN (FIG. 3b) successfully separate out the Zr-rich and Co-Fe binary regions, but it has some unexpected artifact in Fe-rich region. Spectral clustering (FIG. 3c) and agglomerative clustering (FIG. 3d) produce the best results. Both methods pick out the potential phase boundary along Co-Fe binary line, and spectral clustering also identify the phase boundary in Zr-rich region, which are consistent with the other two methods. Both these methods also indicate that the spectra change in regions close to the center of the ternary, and additional possible boundaries that are paralleled to the Co-Fe binary line are observed. This observation was not obvious from attribute mapping or nearest-neighbor distance mapping.

From the NND mapping using 2D images, we know that 2D images contain more information than 1D spectra, and 2D image NND mapping has higher contrast as seen in FIG. 2. The 2D Q-$\chi$ images have 1000*1000 pixels, and a whole dataset will have $10^9$ features, but a generic computer will not be able to handle. Therefore, in order to perform 2D cluster analysis on commodity hardware, the images need first be compressed. Second, the clustering results are highly dependent on the background subtraction method, but a 2D background subtraction method is still unavailable and difficult to develop. Therefore, the NND method is more suitable for 2D image segmentation because it only need 3 patterns at a time and does not require background subtraction.

Summary of segmentation approaches

Within an experimental cycle, all three approaches should be used together as an on-the-fly analysis routine for segmenting XRD patterns, because each of them has some advantages over the others. In the first approach, each attribute represents a particular material characteristic and needs to be defined and constrained carefully by an expert. Therefore, attribute extraction requires the most domain knowledge, but it also provides the most physical insights. Attribute mapping also serves as a partial ground truth for NND and cluster analysis. In the second approach, for each sample in the dataset, the sum of cosine distances of its closest neighbors is recorded. It requires less human effort because the spectra are used directly as inputs without any feature extraction or background subtraction; however, this approach only works when the scanning grids and order are pre-defined. It also does not work well for sparse scanning since the neighbors will be far away from each other, and the NND results will not provide local information anymore. But, because this approach only works with 3 spectra at a time and does not require background subtraction, 2D XRD images, which contain more information than 1D spectra, can be used for segmentation as the data is collected. The NND approach can be used as a first estimate of a phase map. In the last approach, spectral clustering and agglomerative clustering methods produced the best results. The clustering approach requires minimal domain knowledge and often extracts information that is not seen in attribute mapping or NND analysis; they are bench-marks for unsupervised analysis of large datasets. However, they require more computer resources, a careful background subtraction, and cannot work with 2D images on commodity computers.

Expert analysis

Based on the three segmentation approaches, the XRD dataset of Co-Fe-Zr ternary is segmented according to the red lines shown in FIG. 4a. We also know that the spectrum changes more obviously with Zr composition than Co and Fe compositions. Note that in order to confirm the phase boundaries, the spectra from each side of the possible boundary need to be examined carefully. However, the primary focus of this work is to use a set of tools to provide the scientists with preliminary results during data acquisition by providing "real time" suggestions of possible regions for further analysis and investigation. The exact phase mapping of materials-of-interest is out of scope of the current manuscript. For the MG samples studied in this manuscript, it is most important to generate a classification of each observed composition as being amorphous or crystalline with the help of the segmentation results provided by the three approaches. Five samples were then chosen to represent this dataset, at Co:Fe = 1.5 and Zr = 10, 30, 50, 70 and 85 at.%. Their positions are marked by yellow stars in FIG. 4a and their spectra are shown in FIG. 4b. By examining representative diffraction patterns from each segmented region, we were able to quantify the degree of crystallinity of each region. Along the Co-Fe binary for low Zr concentrations, sharp diffraction peaks indicative of crystalline phases are present. The bands in samples along Co-Fe binary region were found to contain a single BCC phase with a FSDP Q position of ~3.1 Å$^{-1}$. Two small peaks centered at about 2.0 Å$^{-1}$ and 3.5 Å$^{-1}$ in Q are from the silicon substrate, which can be observed from all of the spectra. The Fe-Co phase diagram is known to contain a large BCC solid solution region spanning from pure Fe to 80 wt. % Co[29].

When Zr atomic concentration increases above 30 at.% but below 50 at.%, the peak profiles broaden, indicating glass formation from all the three elements. At 70 at.% Zr, the diffraction peaks are broader than typical crystalline peaks but not as broad as those in the glass forming region at lower Zr concentration. Therefore, those samples may be classified as partially-crystalline. Finally at 85 at.% Zr, the samples are observed to be crystalline again. The Zr-rich corner appears to also contain a BCC phase[30] with the primary peak located at 2.5 Å$^{-1}$, both Co and Fe are known to stabilize BCC Zr, although the largest solubility for either element is merely 10 at.%. Expert analysis performed on these 5 samples indicates that many of the compositions in this library are not the ground-state equilibrium structure. The largest change in this ternary is from a crystalline solid-solution near the Co-Fe binary to another crystalline phase towards Zr-rich region, separated by a large region of slowly changing poorly crystalline and amorphous region. A portion of the crystalline region with high Zr centration also has noticeable preferred orientation (texture). Many aspects of this expert analysis are captured in maps produced via all three approaches, but especially so in NND map from the 2D images, agglomerative and spectral clustering.

Here, we employed a series of segmentation approaches to analyze the diffraction dataset. Instead of individually examining more than 1200 individual diffraction patterns, only a handful of patterns (5 in this case study) were needed to understand the morphological changes such as glass-formation and texture in the different regions. By applying this integrated on-the-fly segmentation approach to large XRD dataset, at the end of an experimental cycle, instead of having large quantity of data to analyze for months, scientists are now provided with some preliminary conclusions from the dataset. They can then focus their efforts on the most representative data and work on the dataset more efficiently.

**METHOD**

Thin film alloy deposition

Co-Fe-Zr alloys were co-deposited using single-element targets onto 3-inch Si wafers and form 100 nm thin films. Each of the single element targets are calibrated by controlling the deposition rate at various gun powers and gun tilts and fitted by in-house sputter model software. The ternary system was also synthesized using a deposition rate greater than 0.25 Å/s.

XRD measurement

The as-deposited wafers were studied using synchrotron high throughput x-ray diffraction at Beamline 1-5 at Stanford Synchrotron Radiation Lightsource. The wafers were tilted at a shallow angle (3 degrees) to avoid most of the diffraction from the silicon substrate. The beam size measured directly on the tilted stage is about 3 mm by 0.3 mm. The maximum variation in composition over the beam is 2.94 at.% Co, 1.84 at.% Fe, and 3.36 at.% Zr. The measured spots on the wafer have a spacing of 3 mm. Simultaneously with XRD data collection, a fluorescence detector (Vortex) was be used to track the compositions across the samples allowed by the beam energy (12.7 keV). The Co and Fe signals were recorded in this case. The XRF signals from Co and Fe channels were used to optimize the in-house sputter model software, which was used to determine the compositions of Co, Fe, and Zr at each point. The sputter model software compositions were confirmed via WDS and found to have an average error of 0.59 at.% Co, 1.72 at.% V, and 1.61 at.% Zr.

XRD data processing

The 2D XRD patterns collected by the 2D MARCCD detector were first cleaned using a mean filter to remove the zingers. By using a standard $LaB_6$ material as calibrate, we extracted the geometric parameters of the detector including the sample to detector distance, beam position relative to the detector, the tilting and rotation angles of the detector. The geometric parameters were used to re-mesh the raw 2D XRD images into Q-χ calibrated images and to average into 1D spectra.

**ADDITIONAL INFORMATION**

All the data are organized database and uploaded to Citrine.io (https://citrination.com/datasets/153238/show_search) and Materials website at NIST (http://hdl.handle.net/11256/945). The source code used to generate all the plots can be downloaded at https://github.com/fang-ren/Unsupervised_data_analysis_CoFeZr.

**ACKNOWLEDGEMENT**

This work was supported by the Advanced Manufacturing Office of the Department of Energy under FWP-100250. Use of the Stanford Synchrotron Radiation Lightsource, SLAC National Accelerator Laboratory, is supported by the U.S. Department of Energy, Office of Science, Office of Basic Energy Sciences under Contract No. DE-AC02-76SF00515. T. Williams was supported by NSF IGERT Grant #1250052. The authors thank D. Van Campen and T. Dunn for their support when collecting XRD data at SSRL Beamline1-5. The authors would like to thank the South Carolina Center of Economic Excellence for Strategic Approaches to the Generation of Electricity for support in sample synthesis. The authors acknowledge 2016 Machine Learning for Materials Research Workshop sponsored by NIST and University of Maryland for the introduction on unsupervised machine learning. The authors also thank D. Schneider for recommending cluster analysis methods for this work.

**REFERENCES**

1. T.L. Blundell and S. Patel: High-throughput X-ray crystallography for drug discovery. *Curr Opin Pharmacol* **4**, 490-496 (2004).
2. M. Tanaka, Y. Katsuya and A. Yamamoto: A new large radius imaging plate camera for high-resolution and high-throughput synchrotron x-ray powder diffraction by multiexposure method. *Rev Sci Instrum* **79**, 075106 (2008).
3. J.M. Gregoire, D. Dale, A. Kazimirov, F.J. DiSalvo and R.B. van Dover: High energy x-ray diffraction/x-ray fluorescence spectroscopy for high-throughput analysis of composition spread thin films. *Rev Sci Instrum* **80**, 123905 (2009).
4. J.M. Gregoire, D.G. Van Campen, C.E. Miller, R.J.R. Jones, S.K. Suram and A. Mehta: High-throughput synchrotron X-ray diffraction for combinatorial phase mapping. *J Synchrotron Radiat* **21**, 1262-1268 (2014).
5. S.K. Suram, Newhous, P. F., Zhou, L., Van Campen, D. G., Mehta, A., Gregoire, J. M.: High throughput light absorber discovery, Part 2: Establishing structure-band gap energy relationships. *Acs Comb Sci* **18**, 682-688 (2016).
6. Y.P. Deng, Y. Guan, J.D. Fowkes, S.Q. Wen, F.X. Liu, G.M. Phaff, P.K. Liaw, C.T. Liu and P.D. Rack: A combinatorial thin film sputtering approach for synthesizing and characterizing ternary ZrCuAl metallic glasses. *Intermetallics* **15**, 1208-1216 (2007).
7. M.L. Green, I. Takeuchi and J.R. Hattrick-Simpers: Applications of high throughput (combinatorial) methodologies to electronic, magnetic, optical, and energy-related materials. *J Appl Phys* **113**, 231101 (2013).
8. S.Y. Ding, Y.H. Liu, Y.L. Li, Z. Liu, S. Sohn, F.J. Walker and J. Schroers: Combinatorial development of bulk metallic glasses. *Nat Mater* **13**, 494-500 (2014).
9. M.D. Demetriou, M.E. Launey, G. Garrett, J.P. Schramm, D.C. Hofmann, W.L. Johnson and R.O. Ritchie: A damage-tolerant glass. *Nat Mater* **10**, 123-128 (2011).
10. Y.H. Liu, G. Wang, R.J. Wang, D.Q. Zhao, M.X. Pan and W.H. Wang: Super plastic bulk metallic glasses at room temperature. *Science* **315**, 1385-1388 (2007).
11. J.P. Chu, J.S.C. Jang, J.C. Huang, H.S. Chou, Y. Yang, J.C. Ye, Y.C. Wang, J.W. Lee, F.X. Liu, P.K. Liaw, Y.C. Chen, C.M. Lee, C.L. Li and C. Rullyani: Thin film metallic glasses: Unique properties and potential applications. *Thin Solid Films* **520**, 5097-5122 (2012).
12. D.B. Miracle and O.N. Senkov: A critical review of high entropy alloys and related concepts. *Acta Mater* **122**, 448-511 (2017).
13. D.B. Miracle: Critical Assessment 14: High entropy alloys and their development as structural materials. *Mater Sci Tech-Lond* **31**, 1142-1147 (2015).
14. W.H. Wang, C. Dong and C.H. Shek: Bulk metallic glasses. *Mat Sci Eng R* **44**, 45-89 (2004).
15. M.H. Tsai and J.W. Yeh: High-Entropy Alloys: A Critical Review. *Mater Res Lett* **2**, 107-123 (2014).
16. X. Yang and Y. Zhang: Prediction of high-entropy stabilized solid-solution in multi-component alloys. *Mater Chem Phys* **132**, 233-238 (2012).
17. E. Perim, D. Lee, Y.H. Liu, C. Toher, P. Gong, Y.L. Li, W.N. Simmons, O. Levy, J.J. Vlassak, J. Schroers and S. Curtarolo: Spectral descriptors for bulk metallic glasses based on the thermodynamics of competing crystalline phases. *Nat Commun* **7**, 12315 (2016).

18. K.J. Laws, D.B. Miracle and M. Ferry: A predictive structural model for bulk metallic glasses. *Nat Commun* **6**, 8123 (2015).
19. K. Zhang, B. Dice, Y.H. Liu, J. Schroers, M.D. Shattuck and C.S. O'Hern: On the origin of multi-component bulk metallic glasses: Atomic size mismatches and de-mixing. *J Chem Phys* **143**, 054501 (2015).
20. F. Ren, R. Pandolfi, D. Van Campen, A. Hexemer and A. Mehta: On-the-fly Data Assessment for High Throughput X-ray Diffraction Measurement. *Acs Comb Sci* **19**, 377-385 (2017).
21. The Gaussian-Lorentzian Sum, Product, and Convolution (Voigt) Functions Used in Peak Fitting XPS Narrow Scans, and an Introduction to the Impulse Function, in Vacuum Technology & Coating, edited by M. R. Linford (Weston, CT, 2014), pp. 2-9.
22. A.L. Patterson: The Scherrer formula for x-ray particle size determination. *Phys Rev* **56**, 978-982 (1939).
23. Y. Iwasaki, A.G. Kusne and I. Takeuchi: Comparison of dissimilarity measures for cluster analysis of X-ray diffraction data from combinatorial libraries. *npj Computational Materials* **3**, 4 (2017).
24. A.G. Kusne, T.R. Gao, A. Mehta, L.Q. Ke, M.C. Nguyen, K.M. Ho, V. Antropov, C.Z. Wang, M.J. Kramer, C. Long and I. Takeuchi: On-the-fly machine-learning for high-throughput experiments: search for rare-earth-free permanent magnets. *Sci Rep-Uk* **4**, 6367 (2014).
25. A.G. Kusne, D. Keller, A. Anderson, A. Zaban and I. Takeuchi: High-throughput determination of structural phase diagram and constituent phases using GRENDEL. *Nanotechnology* **26**, 444002 (2015).
26. H.S. Stein, S. Jiao and A. Ludwig: Expediting Combinatorial Data Set Analysis by Combining Human and Algorithmic Analysis. *Acs Comb Sci* **19**, 1-8 (2017).
27. S.K. Suram, Y.X. Xue, J.W. Bai, R. Le Bras, B. Rappazzo, R. Bernstein, J. Bjorck, L. Zhou, R.B. van Dover, C.P. Gomes and J.M. Gregoire: Automated Phase Mapping with AgileFD and its Application to Light Absorber Discovery in the V-Mn-Nb Oxide System. *Acs Comb Sci* **19**, 37-46 (2017).
28. F. Pedregosa, G. Varoquaux, A. Gramfort, V. Michel, B. Thirion, O. Grisel, M. Blondel, P. Prettenhofer, R. Weiss, V. Dubourg, J. Vanderplas, A. Passos, D. Cournapeau, M. Brucher, M. Perrot and E. Duchesnay: Scikit-learn: Machine Learning in Python. *Journal of Machine Learning Research* **12**, 2825-2830 (2011).
29. D. Hunter, W. Osborn, K. Wang, N. Kazantseva, J. Hattrick-Simpers, R. Suchoski, R. Takahashi, M.L. Young, A. Mehta, L.A. Bendersky, S.E. Lofland, M. Wuttig and I. Takeuchi: Giant magnetostriction in annealed $Co_{1-x}Fe_x$ thin-films. *Nat Commun* **2**, 518 (2011).
30. A. Jain, S.P. Ong, G. Hautier, W. Chen, W.D. Richards, S. Dacek, S. Cholia, D. Gunter, D. Skinner, G. Ceder and K.A. Persson: Commentary: The Materials Project: A materials genome approach to accelerating materials innovation. *Apl Mater* **1**, 011002 (2013).

**LIST OF CAPTIONS**

FIG. 1. Attribute mapping of Co-Fe-Zr ternary. (a) Imax/Iave. (b) square sum of texture. (c) number of peaks. (d) intensity of FSDP. (e) FWHM of FSDP. (f) position of FSDP.

FIG. 2. NND histograms of Co-Fe-Zr ternary based on (a) XRD 1D spectra and (b) XRD 2D Q-$\chi$ images. Inset: NND map with magnified views of Zr-rich region.

FIG. 3. Clustering of 1D spectra of XRD results of Co-Fe-Zr. (a) K-medoids (n_clusters = 7) using 1D spectra. (b) DBSCAN (eps = 0.0009, min_sample = 5) using 1D spectra. (c) Spectral clustering (n_ clusters = 6) using 1D spectra. (d) Agglomerative clustering (n_ clusters = 6, linkage = 'average') using 1D spectra.

FIG. 4.(a) Phase boundary in Co-Fe-Zr ternary and 5 representatives; (b) XRD spectra for the 5 representatives.